\documentclass[conference]{IEEEtran}
\IEEEoverridecommandlockouts
\usepackage{cite}
\usepackage{amsmath,amssymb,amsfonts}
\usepackage{algorithmic}
\usepackage{graphicx}
\usepackage{textcomp}
\usepackage{xcolor}
\usepackage{multirow}
\usepackage{makecell}
\usepackage{booktabs}
\def\BibTeX{{\rm B\kern-.05em{\sc i\kern-.025em b}\kern-.08em
    T\kern-.1667em\lower.7ex\hbox{E}\kern-.125emX}}
\begin{document}

\title{Demystifying Bitcoin Address Behavior via Graph Neural Networks}

\author{
    Zhengjie Huang\IEEEauthorrefmark{2}, 
	Yunyang Huang\IEEEauthorrefmark{3}, 
	Peng Qian\IEEEauthorrefmark{2}\IEEEauthorrefmark{1}, 
	Jianhai Chen\IEEEauthorrefmark{2},
	Qinming He\IEEEauthorrefmark{2}\IEEEauthorrefmark{1}\\
	\IEEEauthorrefmark{2}College of Computer Science and Technology, Zhejiang University, Hangzhou, China\\
	\IEEEauthorrefmark{3}School of Information and Communication Engineering, UESTC, Chengdu, China\\
	\thanks{\IEEEauthorrefmark{1} Peng Qian and Qinming He are the corresponding authors.}
	zj.h@zju.edu.cn, \{yunyanghuang96, messi.qp711\}@gmail.com\\
	\{chenjh919, hqm\}@zju.edu.cn
}

\maketitle

\begin{abstract}
Bitcoin is one of the decentralized cryptocurrencies powered by a peer-to-peer blockchain network. Parties who trade in the bitcoin network are not required to disclose any personal information. Such property of anonymity, however, precipitates potential malicious transactions to a certain extent. Indeed, various illegal activities such as money laundering, dark network trading, and gambling in the bitcoin network are nothing new now. While a proliferation of work has been developed to identify malicious bitcoin transactions, the behavior analysis and classification of bitcoin addresses are largely overlooked by existing tools. In this paper, we propose {BAClassifier}, a tool that can automatically {classify bitcoin addresses based on their behaviors}. Technically, we come up with the following three key designs. First, we consider casting the transactions of the bitcoin address into an address graph structure, of which we introduce a graph node compression technique and a graph structure augmentation method to characterize a unified graph representation. Furthermore, we leverage a graph feature network to learn the graph representations of each address and generate the graph embeddings. Finally, we aggregate all graph embeddings of an address into the address-level representation, and engage in a classification model to give the address behavior classification. As a side contribution, we construct and release a large-scale annotated dataset that consists of over 2 million real-world bitcoin addresses and concerns 4 types of address behaviors. Experimental results demonstrate that our proposed framework outperforms state-of-the-art bitcoin address classifiers and existing classification models, where the precision and F1-score are 96\% and 95\%, respectively. Our implementation and dataset are released, hoping to inspire others.
\end{abstract}


\section{Introduction}

Bitcoin is a kind of cryptocurrency asset that {{originates}} as a concept described in~\cite{nakamoto2008bitcoin}. It utilizes a distributed ledger technology called \emph{blockchain} to store and transfer assets in a peer-to-peer fashion {{without the involvement of trusted third parties}}. Bitcoin is widely regarded as the foundation of crypto assets, and {{several types of digital currencies have been generated and broken off from the Bitcoin blockchain}}~\cite{monrat2019survey}. Bitcoin now is comprised of over 33,000 active nodes, enjoying a market capitalization greater than \$685 billion.

In essence, blockchain is a decentralized protocol to securely store bitcoin transaction data. A blockchain is a decentralized ledger maintained by global bookkeeping nodes (known as miners) following a common consensus mechanism~\cite{leonardos2020presto, harvey2014bitcoin, gupta2021rcc,gupta2021fault,gupta2019proof,gupta2020resilientdb,qian2019digital}. {{The ledger, which is fully replicated over all bookkeeping nodes worldwide, enables the immutability of transactions}}, endowing blockchain with its tamper-proof nature. {{Additionally, in the bitcoin network, the identity of the account owner is unknown to everyone. Put differently, we do not know whom addresses in the bitcoin network are associated with in the real world.}} Individuals are able to transfer money through bitcoin at their leisure without the need to verify any identifying information. The process is similar to a bank transfer without being connected to any bank card. However, such a trait has given birth to various illegal bitcoin tradings such as money laundering~\cite{fanusie2018bitcoin, wu2021detecting,weber2019anti}. Black market and dark network are becoming the most popular platforms for illegal bitcoin trading~\cite{janze2017cryptocurrencies,christin2013traveling,aldridge2016hidden,yin2017first}.

A transaction in the bitcoin network is the process of transferring bitcoins from one bitcoin address to another. A bitcoin address is derived from the asymmetric key system in cryptography and is a 26-bit to 34-bit string of letters and numbers. 
{{Here, we present a simplified bitcoin transfer scenario as: 1) First, Alice creates a transaction, where she digitally signs with her private key to state her ownership of the bitcoins that will be transferred. 2) Alice issues a locking “declaration” that only Bob’s signature can take out the bitcoins. The transaction is then transmitted to the bitcoin network and becomes part of a block in the bitcoin network. 3) Finally, Bob is capable of utilizing his signature to unlock this transaction, gaining the bitcoins transferred from Alice.}}

\begin{figure}
    \centering
    \includegraphics[width=1\linewidth]{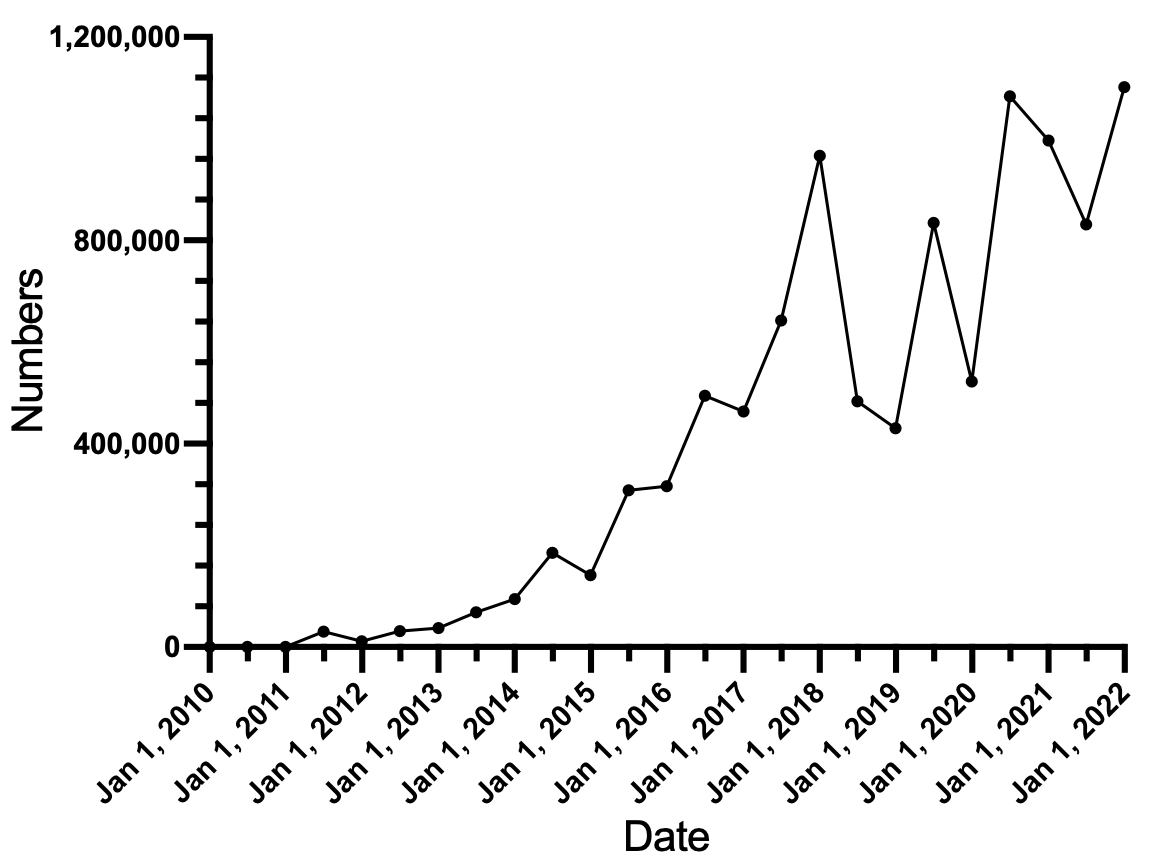}
    \caption{Illustration of the number of active bitcoin addresses. With an increase in the popularity of bitcoin and its price, the number of active bitcoin addresses has exceeded 1.1 million as of January 2022.}
    \label{fig:NAA}
\end{figure}

So far, the behavior analysis of bitcoin addresses has attracted extensive attention. A critical research direction is to collect and analyze bitcoin address transactions in order to ascertain the sort of entity to which the bitcoin address belongs. The possible types of entities include exchanges, mining pools, gambling, and so on. As a consequence, an increasing number of illegal bitcoin transaction activities are emerging. To strengthen the monitoring and tracking of illegal bitcoin transactions, it is extremely important to implement the classification for large-scale bitcoin addresses~\cite{kuzuno2017blockchain}. 
As shown in Fig.~\ref{fig:NAA}, the number of active bitcoin addresses has grown tenfold over the last decade. With the continuous growth of bitcoin addresses, an accurate yet efficient method for address classification is much coveted.

Existing methods for bitcoin address classification can be roughly cast into two categories, \emph{i.e.,} off-chain information-based method and on-chain behavior-based method. The off-chain information-based method focuses on tagging addresses by gathering real-world data from address users. For example, Ermilov \emph{et al.}~\cite{ermilov2017automatic} crawl the user profiles from relevant forums (\emph{e.g.,} Bitcointalk.com, Twitter, and Reddit) and darknet markets (\emph{e.g.,} Silkroad, Hub Marketplace, and Alphabay) to obtain the association between some bitcoin addresses and users, and then analyze the behavior of other addresses using {{a clustering method}}. Kang \emph{et al.}~\cite{kang2020anonymization} obtain the IP address by receiving bitcoin protocol message packets, and then use static analysis to infer other addresses which the current IP address may be associated with, realizing the mapping between bitcoin addresses and IP addresses. Unfortunately, these solutions are highly dependent on off-chain collected information and human experience, and hence cannot be used for all bitcoin addresses. Additionally, there are usually many mistakes in off-chain information, which might lead to low accuracy of the address behavior analysis. 

The on-chain behavior-based method concentrates on extracting transaction characteristics of bitcoin addresses and analyzing their behavior with the guidance of machine learning. For instance,~\cite{lee2020machine,burks2017bitcoin,lin2019evaluation,toyoda2018multi,zambre2013analysis,kanemura2019identification} directly extract transaction features from bitcoin addresses and then feed them into models, e.g., Random Forest, SVM, and LightGBM. 
Such methods do not need to rely on off-chain information, thus avoiding potential problems caused by missing off-chain information. However, they also suffer from inherent drawbacks. On one hand, direct feature extraction from transactions in the addresses may lead to large deviations. On the other hand, the traditional classification model has difficulties in utilizing the temporal feature and topology of transactions in the addresses. These problems impose a significant impact on the accuracy of address behavior analysis, which motivates us to design a novel and effective address behavior analysis model.

In this paper, we present BAClassifier, a fully automatic framework for bitcoin address classification. In particular, BAClassifier consists of three key components. \emph{(1) Address Graph Construction.} {{For each given bitcoin address, BAClassifier constructs a chronological transaction graph that reflects the behavior of that address}}. Specifically, we engage in a graph node compression technique and a graph structure augmentation method to transfer the transactions of each bitcoin address into a unified graph representation. \emph{(2) Graph Representation Learning.} BAClassifier utilizes graph neural networks to learn the graph representation of each address and generate the graph embeddings. \emph{(3) Address Classification.} BAClassifier aggregates all graph embeddings of each address into the address-level representations, and engages in a classification model to give the final address behavior classification.


{{
In summary, we present the following \textbf{key contributions}: 
\begin{itemize}
\item We propose BAClassifier, a tool that can automatically classify bitcoin address behaviors. Particularly, we investigate a scalable and generic manner of analyzing bitcoin address behaviors using graph neural networks.
\item Within this system, we come up with three key components, \emph{i.e.,} address graph construction, graph representation learning, and address classification. Specifically, we introduce graph node compression and graph structure augmentation techniques to translate bitcoin address transactions into unified graph representations. Furthermore, we adopt graph feature networks to extract address graph features and select the combination of LSTM+MLP as the behavior classification model.
\item We construct a large-scale labeled dataset that consists of over 2 million bitcoin addresses as well as their transactions and concerns 4 types of address behaviors, which can serve as a benchmark for evaluating bitcoin address classification methods.
\item Extensive experiments on the collected dataset show that our proposed system is indeed useful in identifying bitcoin address behaviors. BAClassifier surpasses state-of-the-art address classifiers and overall provides interesting insights. We have released our code and dataset at https://github.com/AwesomeHuang/BAClassifier, hoping to facilitate future research.
\end{itemize}
}}

\section{Background}


\subsection{Bitcoin UTXO Model}
{{The Bitcoin transaction model is distinct from the conventional account-based model}}. It lacks a real field to record the balance of users, thus relying on UTXOs, each of which is associated with a bitcoin address. UTXO is short for Unspent Transaction Output. One bitcoin transaction usually consists of multiple input addresses and output addresses. A bitcoin wallet is a client for users to manage bitcoin addresses. {{Different from traditional wallets, Bitcoin wallets only record the public and private keys of all UTXOs of the user, and do not directly hold Bitcoins or participate in transactions. Once a transaction is launched, the bitcoin wallet will look through all the available UTXOs and determine the right amount to pay. Undoubtedly, the fragmentation of bitcoin transactions will result in a fast increase in the number of UTXO addresses. To solve this problem, the Bitcoin change mechanism was designed.

When a transaction occurs, the bitcoin wallet will zero off the balance in the original address, and transfer any leftover funds to a new address. Naturally, the address that receives the change could be set to the original address of the originating transaction. Then, the bitcoin wallet automatically generates a new address to receive the change after the transferred amount and the fee are deducted.}}  Such a mechanism safeguards the user’s privacy by ensuring that no one other than the user could know which address is the change address and which is the receipt address. However, all these facts make the analysis of bitcoin address behavior more difficult and challenging than that of traditional account models.

{{
\subsection{Machine Learning Technique}
Recent years have witnessed a dramatic rise in the popularity of machine learning techniques. After investigating a variety of machine-learning applications, we found that machine-learning tasks could be roughly classified into the following three categories, classification, clustering, and regression. One of the most common tasks is classification, and our bitcoin address behavior detection task falls into this category. Algorithms that are used for classification tasks are usually linear classification (\emph{e.g.,} logistic regression~\cite{wright1995logistic}, support vector machines~\cite{hearst1998support}), Bayesian classification (\emph{e.g.,} Bernoulli Naive Bayesian, Gaussian Naive Bayesian~\cite{leung2007naive}), decision trees (\emph{e.g.,} ID3~\cite{peng2009implementation}, CART~\cite{lewis2000introduction}), and ensemble learning (\emph{e.g.,} GBDT~\cite{friedman2001greedy}, XGBoost~\cite{chen2016xgboost}). All of these algorithms are able to achieve classification tasks based on the features of input data. However, machine learning-based methods still have some shortcomings, such as the inability to learn the correlation between features, which thus cannot be applied to the classification of images, videos, and graph data. It is worth mentioning that the emergence of deep learning has solved these problems. As a new branch of machine learning techniques, deep learning models provide new solutions for computer vision (\emph{e.g.,} CNN~\cite{wang2019development}), natural language processing (\emph{e.g.,} BERT~\cite{devlin2018bert}, Transformer~\cite{vaswani2017attention}), social networks (\emph{e.g.,} DeepWalk~\cite{perozzi2014deepwalk}, GNN~\cite{henaff2015deep}), adversarial learning (\emph{e.g.,} GAN~\cite{pan2019recent}), and other fields.

\begin{figure*}
    \centering
    \includegraphics[width=18.6cm]{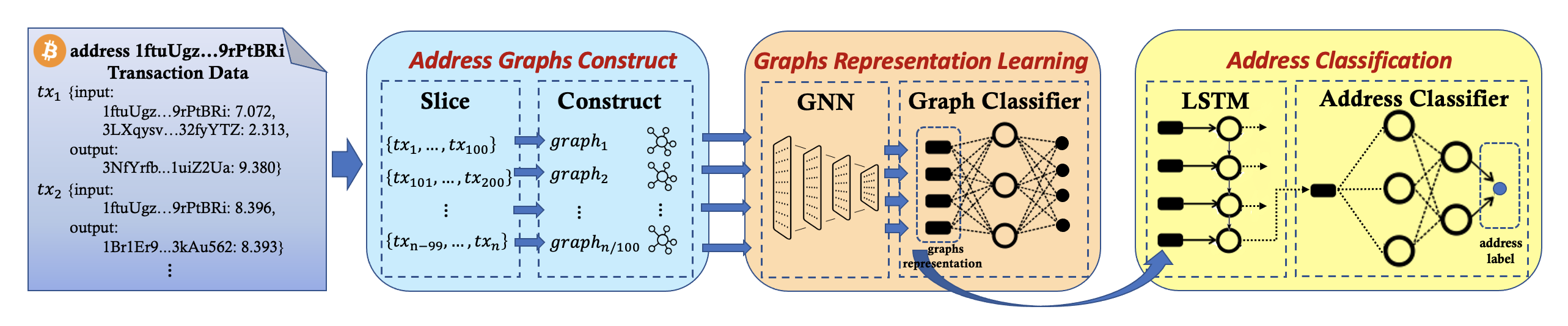}
    \caption{The high-level pipeline of the proposed BAClassifier framework, which consists of three major modules, namely address graph construction, graph representation learning, and address classification.}
    \label{fig:AS}
\end{figure*}

\subsection{Graph Neural Network}
While traditional neural network-based methods have been successfully used to extract features from Euclidean data, in many practical application scenarios where data are generated from a non-Euclidean space, their performance is far from satisfactory~\cite{henaff2015deep}. In recent years, graph neural networks (GNNs) that can handle non-Euclidean structured data have been investigated extensively in various fields such as graph classification~\cite{shen2021identity,errica2019fair,wang2020graphcrop,le2021parameterized,zhuang2020smart,liu2021smart}, pattern recognition~\cite{li2020dynamic,cardot2002graph,kajla2021graph,shi2020point,yun2021instance,gong2021mmpoint}, and data mining~\cite{pan2019learning,yeh2019novel,gundem2010intogen,yang2019aligraph}. Graph neural networks take features of edges and nodes as inputs and generate outputs tailored to specific tasks. Existing graph neural networks can be roughly cast into two categories. One line of work generalizes well-established neural networks like CNNs to work on graph-structured data. Another line of work inherits ideas from recurrent GNNs and adopts information propagation to define graph convolutions. 

Graph neural networks are usually used to handle the following three tasks. (1) Node-level tasks, where the output is used for node regression and classification. GraphSage~\cite{hamilton2017inductive}, GAT~\cite{velivckovic2017graph}, N-GCN~\cite{abu2020n}, VR-GCN~\cite{ye2019vectorized}, PinSage~\cite{ying2018graph}, and related models are often used to solve the node-level tasks. (2) Edge-level tasks, where the output is related to edge classification~\cite{gong2019exploiting} and link prediction~\cite{hu2020heterogeneous, cen2019representation}. As with~\cite{schlichtkrull2018modeling}, the graph neural network used to do edge prediction consists of two main components: an encoding component for node representation, and a decoding component for generating edge relationships based on the representations of the nodes at both ends of the edge. (3) Graph-level tasks, where the output is related to graph classification tasks, such as graph representation. Models such as Graph Convolutional Networks~\cite{DBLP:conf/iclr/KipfW17}, DiffPool~\cite{ying2018hierarchical}, SAGPool~\cite{lee2019self}, EigenPool~\cite{ma2019graph}, GIN~\cite{xu2018powerful}, Graph Feature Network~\cite{chen2019powerful} are usually adopted to deal with graph-level tasks. In our work, each address in the bitcoin network can be treated as a node. Different nodes are linked by transactions, forming the structure of transaction graphs. Such a fact motivates us to consider using graph neural networks to address the analysis of bitcoin address behavior.
}}

\section{Our Method}
\label{Method Architecture}

\textbf{Problem Formulation.}\quad Given a bitcoin address, we are interested in developing a fully automated approach that can analyze the address behavior and determine the specific address type. {{Formally, given a list of bitcoin addresses $\mathbb{A}=\{ad_1, ad_2, ..., ad_n\}$, we aim to predict the labels $\mathcal{Y}=\{y_1, y_2, ..., y_n\}$, for each $y_i$ $\in$ $\mathcal{Y}$, $i$ $\in$ $\{1, 2, ..., n\}$, $y_i$ $\in$ $\{1, 2, ..., t\}$. Here, $n$ represents the length of list $\mathbb{A}$ while $t$ denotes the number of defined address behavior types.}}

For each address $ad_i \in \mathbb{A}$, we gather all the transactions related to $ad_i$. Since the number of transactions of each address varies, we define the transactions {{$T_i = \{tx_{i1}, tx_{i2}, ..., tx_{im}\}$ related to $ad_i$, where $im$ denotes the number of transactions of $ad_i$.}} As such, we can obtain the transaction list $\mathbb{T} = \{T_1, T_2, ..., T_n\}$ corresponding to the address list $\mathbb{A}$.

In this work, we consider using graph neural networks to analyze the bitcoin address behaviors, which is actually a multi-classification task. Before diving into the method, we first introduce two key issues that may directly influence the performance of the address classification.

\emph{(1) How to construct a unified structure of address transaction graphs? } {{Different bitcoin addresses have a distinct number of transactions. In particular, the number of input and output addresses of different transactions varies greatly, ranging from a few hundred to many thousands. The scales of graphs constructed with different addresses will be quite different. On one hand, graphs with different sizes and structures will hinder the model learning of graph neural networks. On the other hand, when graphs of different scales are directly transferred to the graph model, it is difficult to accurately extract graph features, thus affecting the accuracy of classification tasks. Therefore, it is critical to design corresponding methods to maintain a unified structure of extract address transaction graphs.}}


\emph{(2) How to pick out an appropriate graph classification model? } The bitcoin network organizes transactions with the UTXO model, resulting in a large number of addresses in the graph data. This requires the model to improve the processing speed of graph data while maintaining the accuracy of feature expression. Additionally, bitcoin addresses have a temporal relationship between transactions. There are timestamps in transactions, and each transaction is associated with a specific block. It is still challenging for the graph neural network to handle graphs with temporal characteristics.

\textbf{Method Overview.}\quad The overall architecture of BAClassifier is outlined in Fig.~\ref{fig:AS}. Generally, BAClassifier consists of three key components: 
\begin{itemize}
\item \emph{Address Graph Construction:} For each given bitcoin address, BAClassifier will construct a chronological transaction graph that reflects the behavior of that address.
\item \emph{Graph Representation Learning:} BAClassifier utilizes a graph neural network to learn the graph representation of each address and generate the graph embeddings.
\item \emph{Address Classification:} BAClassifier aggregates all graph embeddings to produce the address-level representations, and confirms the classification model to output the predictions of address classifications.
\end{itemize}
In what follows, we will elaborate on the details of these components one by one.



\subsection{Address Graph Construction}
\label{Graph Construction}

{{The first step in BAClassifier is to transfer the address transactions into graph structures.}} To obtain a unified address graph, we have to solve three key problems. 
(1) Different bitcoin addresses have a distinct number of transactions, thus yielding different sizes of graphs. Moreover, the transactions of one address are performed in temporal order. Therefore, we must guarantee that the generated graphs have a unified structure while still preserving the temporal order of transactions. 
(2) There is a significant disparity in the number of addresses involved in various transactions. For instance, a transaction issued by certain exchanges may only have several associated addresses, while a transaction generated by a mining pool may have thousands of associated addresses. Hence, the graph size must be limited. 
(3) Since bitcoin transactions provide an insufficient amount of information, we thus need to go deeper into the global graph structural feature to elicit further information. 
To tackle the above problems, BAClassifier incorporates three key modules into the address graph construction, \emph{namely} original graph extraction, graph node compression, and graph structure augmentation.

\subsubsection{Original Graph Extraction}
For address $ad_i$, we construct the original graph list $\mathbb{G} = \{G_{ad_i}^{1}, G_{ad_i}^{2}, ..., G_{ad_i}^{g\_num }\}$. For each graph $G_{ad_i}^{j}, j \in \{1,2, ..., g\_num\}$, it can be represented as $G_{ad_i}^{j} = (V_{ad_i}^{j}, E_{ad_i}^{j})$. For convenience, we simplify $G_{ad_i}^{j} = (V_{ad_i}^{j}, E_{ad_i}^{j})$ as $G = (V, E)$. $G$ is a heterogeneous graph, which has two types of nodes, \emph{i.e.,} address node $v^{addr} \in V$ and transaction node $v^{tx} \in V$. {{Edge $e_{ij} = (v_i^{addr},v_j^{tx} ) \in E$ in $G$ denotes that an address node $v^{addr}$ points to a transaction node $v^{tx}$, implying that address $ad_i$ is involved in the transaction $tx_{ij}$.}} Notably, we use $value(e_{ij})$ to represent the amount transferred by the address $ad_i$.

We split all transactions within an address into several groups according to the timestamp. The number of transactions in each graph is fixed. In practice, we select 100 transactions as the slicing unit, {{which implies that one address could construct $g\_num = \left \lceil im/100 \right \rceil $ graphs, and each graph contains the information from the 100 transactions. It is worth noting that the final graph with less than 100 transactions will be retained.}}

\subsubsection{Graph Node Compression}
\begin{figure}
    \centering
    \includegraphics[width=1\linewidth]{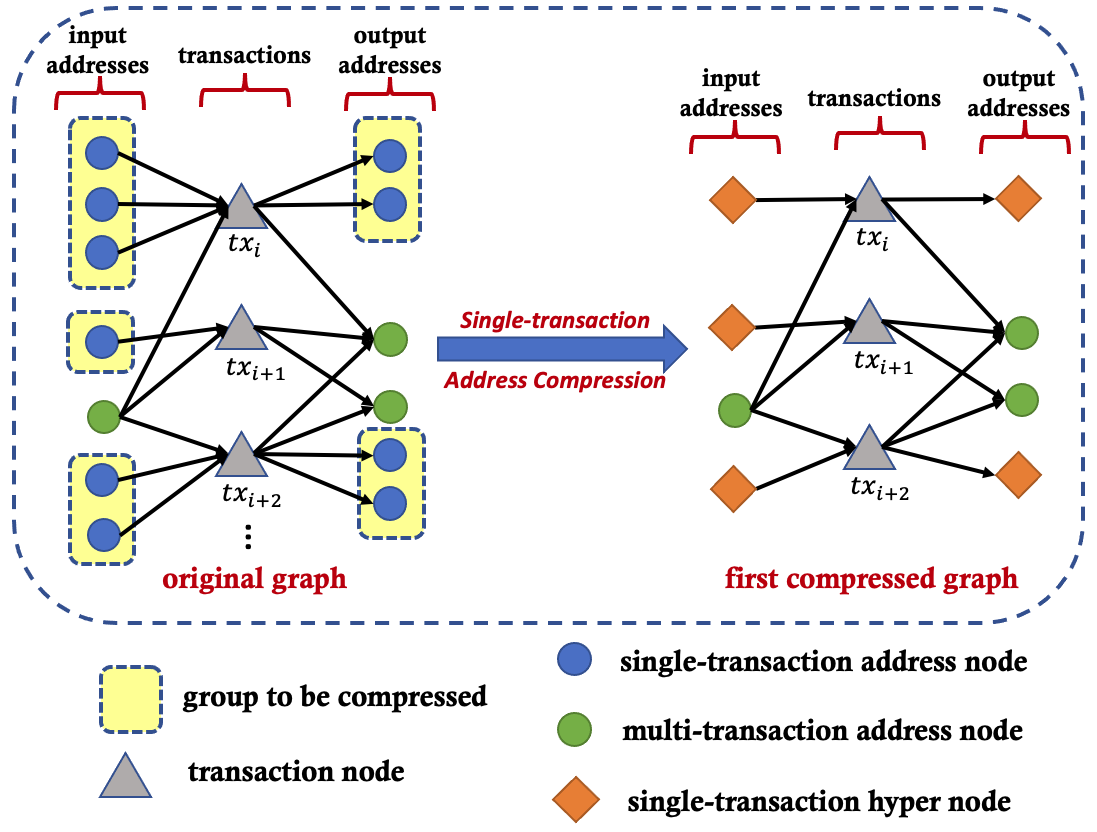}
    \caption{Illustration of the single-transaction address compression. The proposed SFE approach merged addresses that are connected to one transaction into a single-transaction hyper node. }
    \label{fig:SAC}
\end{figure}

\begin{figure}
    \centering
    \includegraphics[width=1\linewidth]{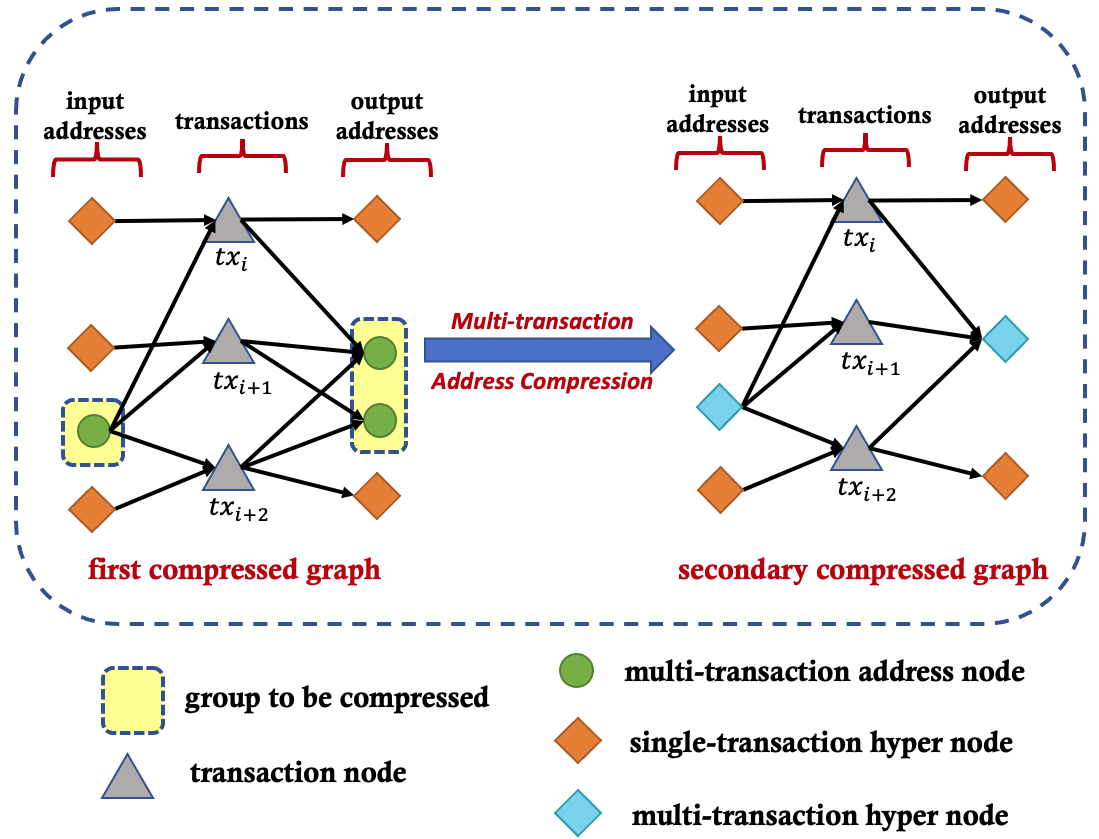}
    \caption{Illustration of the multi-transaction address compression. The proposed SFE approach merged addresses with similar connectivity relationships into a multi-transaction hyper node.}
    \label{fig:MAC}
\end{figure}

{{On one hand, different bitcoin addresses will yield distinct scales of address transaction graphs, hindering the training of graph neural networks. Moreover, most graph neural networks are inherently flat when propagating information, ignoring key features of certain nodes~\cite{liu2021combining}. On the other hand, large-scale address transaction graphs seriously affect the training efficiency of a model. As a result, we consider proposing a graph normalization method to deal with originally extracted address graphs. Empirically, we observe that there are two different types of nodes in original address graphs (\emph{i.e.,} address nodes and transaction nodes), of which address nodes account for the vast majority. In addition, we found out that many transactions issued by some address nodes are not beneficial to address behavior detection, and only partial transactions reflect the real behavior features of those addresses. For example, addresses that create transactions with no value are useless for detecting their behaviors. Motivated by this, we propose a node compression technique, which merges address nodes while preserving key characteristics of each address. We have illustrated the single-transaction address compression and the multiple-transaction address compression in Fig.~\ref{fig:SAC} and Fig.~\ref{fig:MAC}, respectively.}}

Specifically, we classify address nodes into two categories. One is a single-transaction address, which contains only one transaction, \emph{i.e.,} the out-degree of the node is 1. The other is a multi-transaction address, which contains multiple transactions, \emph{i.e.,} the out-degree of the node is greater than 1. To compress address nodes, we introduce a two-stage compression algorithm, which consists of \emph{single-transaction address compression} and \emph{multi-transaction address compression}.

Before introducing the two-stage compression algorithm, we first introduce the \emph{statistical feature extraction (SFE)} method, which is the feature extraction technique to deal with the characteristics of the merged node.

\textbf{Statistical Feature Extraction.}\quad We design the statistical feature extraction method for capturing the transfer feature of each transaction. Since the majority of the available information in a bitcoin transaction is the transferred amount, we utilize SFE to analyze the statistical features of the transferred amount in the address node which will be merged. Given $N$ address nodes, we can obtain the aggregated feature of the final merged address node $v_m^{addr}$ by
\begin{equation}
\textit{feature}(v_m^{addr}) = \textit{SFE}(\sum_{k \in N}^{} \sum_{j \in V}^{} \textit{value}(e_{kj}) )
\end{equation}
where $V$ denotes the number of address nodes. SFE takes the transferred amount $value(e_{ij})$ of each address node as the input, and compute the statistical characteristics in the following list.

\begin{itemize}
\item Max, min, sum, mean, and number of the input.
\item Range, mid-range, percentile, variance, and standard deviation of the input.
\item Mean absolute deviation and coefficient of variation of the input.
\item Kurtosis, skewness, and tilt of the input.
\end{itemize}

\textbf{Single-Transaction Address Compression.}\quad The single-transaction address compression is proposed to merge the address nodes that have only a single transaction, reducing the number of address nodes in the graph. Fig.~\ref{fig:SAC} illustrates the specific procedure. For the single-transaction address connected to the same transaction node, we merge these nodes into a hyper node, termed a {single-transaction hyper node}. To retain the input or output value of these addresses, we adopt the statistical feature extraction method to retrieve the transfer feature for these single-transaction hyper nodes, which is the features of the single-transaction hyper node in the graph.
After compressing address nodes, a transaction node can link to a maximum of two single-transaction hyper nodes on the input and output side of the transaction, respectively.

We denote $v_j^{tx}$ as $j$-th transaction node, $v_j^{s\_hyper}$ as the single-transaction hyper node of transaction $t_j$, and $g_j^{single}$ as the node group containing single-transaction address nodes which related to transaction $t_j$. Therefore, we can acquire the feature of the single-transaction hyper node by
\begin{equation}
\textit{feature}(v_j^{s\_hyper}) = \textit{SFE}(\sum_{k \in g_j^{single}}^{} \textit{value}(e_{kj}) )
\end{equation}

\textbf{Multi-Transaction Address Compression.}\quad Regarding the multi-transaction addresses, our investigation uncovered some interesting findings. There are many addresses that occur concurrently in two or more transactions. For example, the mining pool will pay the reward to every address which participated in the mining, resulting in thousands of mining addresses being linked to each transaction of the mining pool address simultaneously. As a result, a large-scale graph will be generated by a significant number of address nodes with similar behaviors.
Due to the fact that these address nodes are associated with multiple transactions, we thus consider using a multi-transaction address compression method to merge such a kind of address nodes, as shown in Fig.~\ref{fig:MAC}.

For convenience, we denote $g^{multi}$ as the group of multi-transaction address nodes in the graph $G$, and denote $v_i^{multi}$ as the $i$-{th} address node in $g^{multi}$. To minimize the size of the address graph, we propose the address correlation calculation framework as follows.
\begin{gather}
S=AA^T \\
M =S*D^{-1} \\
Q=\textit{ReLU}(M-thr_{sim} * I)
\end{gather}
Specifically, $A \in R^{n\times d}$ is the adjacency matrix, where $n$ is the number of nodes in the transaction group $g^{multi}$ and $d$ denotes the number of transactions in graph $G$. $A^T$ stands for the transpose of $A$. The adjacency matrix is used to represent the relationship between multi-transaction address nodes and transactions in $G$. For $a_{ij} \in A$, if $v_i^{multi}$ and $v_j^{tx}$ are connected, $a_{ij} = 1$, otherwise $a_{ij} = 0$. To determine the similarity of two addresses in $g^{multi}$, we count the number of transactions that occurred at both addresses. For $s_{ij} \in S$, $s_{ij}$ is the number of common transactions between two different multi-transaction address nodes. We can find that diagonal elements in $S$ is the total number of transactions connected to each address, namely the degree of the address node. 

Before directly calculating {{node similarity with matrix $S$}}, we first normalize the matrix. Assuming $D$ is degree matrix of $S$, we can get its inverse matrix $D^{-1}$ and use it to normalize the similarity matrix $S$ by calculating $M  = S * D^{-1}$. $M$ is a normalized similarity matrix, and each element in $M$ represents the similarity of two addresses. For instance, $s_{1 1}=10$ indicates that the multi-transaction address node $v_1^{multi}$ is connected to 10 transaction address nodes, and $s_{3 1}$ implies $v_3^{multi}$ and $v_1^{multi}$ are related to 7 identical transaction addresses. As such, $m_{31} = 0.7$  suggests that 70\% of transactions in $v_3^{multi}$ are connected to $v_1^{multi}$. Therefore, we draw the conclusion that the similarity between $v_3^{multi}$ and $v_1^{multi}$ is 0.7.
After obtaining the similarity matrix $M$, for any node $v_i^{multi} \in g^{multi}$, we can figure out which nodes are similar to $v_i^{multi}$ and merge them into $v_i^{multi}$.

To identify similar addresses for every address node $v_i^{multi}$, we set a similarity threshold $\Psi$ and choose those addresses with a similarity larger than $\Psi$. To achieve this goal, we process the normalized similarity matrix $M$ using the ReLU function. We define all-ones matrix $I$, and obtain matrix $Q = \textit{ReLU}(M-\Psi*I)$, for $q_{ij} \in Q$, if $m_{ij}> \Psi$, $q_{ij} \geq 0$, otherwise $q_{ij} = 0$. Let $q_i$ denote the $i$-th row vector in $Q$, $\textit{nonzero}(q_i)$ is the function to calculate the number of non-zero element of $q_i$, and $\textit{addr\_idx}(q_i)$ is the function to return $g_{i}^{sim}\subseteq g^{multi}$, which means the address nodes in $g_{i}^{sim}$ are similar to node $v_i^{multi}$.

We set another threshold $\sigma$ to identify {{addresses that are similar to a large number of other addresses.}} If the number of non-zero element in $q_i$ is the greater than $\sigma$, the node will be retained and merged with similar nodes $g_{i}^{sim}$. We compress the corresponding address into a hyper node $v_j^{m\_hyper}$ and denote $g_{k}^{tx}$ as a group of transaction nodes that connected to the address $ad_k$. Finally, we adopt the statistical feature extraction method to obtain the feature of the multi-transaction hyper node $\textit{feature}(v_j^{m\_hyper})$.
\begin{gather}
g_{i}^{sim} = \textit{add\_idx}(q_i), if\ \textit{nonzero}(q_i) > \Psi \\
\textit{attribute}(v_i^{m\_hyper}) = \textit{SFE}(\sum_{k \in g_i^{sim}}^{} \sum_{j \in g_{k}^{tx}}^{} \textit{value}(e_{kj}) )
\end{gather}

\subsubsection{Graph Structure Augmentation}
By compressing address nodes, we can obtain a set of compressed address graphs. We further propose a network centrality strategy to enhance the structural information of the nodes in the graph. Network centrality includes degree centrality, betweenness centrality, closeness centrality, and PageRank centrality. We incorporate such features of network centrality into the graph nodes. 


\textbf{Degree centrality} is the degree of a node. Degree centrality of a node $v_i$ is defined as:
\begin{equation}
\textit{C}_D(v_i) = \textit{degree}(v_i)
\end{equation}

\textbf{Closeness centrality} is used to quantify the difficulty of a node to reach other nodes~\cite{sankar2019meta}. It is calculated as the reciprocal of the average distance to all other nodes.
\begin{equation}
\textit{C}_C(v_i) = \frac{\left | V \right | -1}{ {\textstyle \sum_{v_i\ne t\in V}^{}d_{v_{i}t}} }
\end{equation}
where $d_{v_{i}t}$ denotes the shortest path between node $t$ and $v_i$, and $\left | V \right |$ is the number of nodes in the graph.

\textbf{Betweenness centrality} measures the importance of nodes in terms of their connectivity to other nodes via the shortest paths~\cite{fan2019learning}. Betweenness centrality of a node $v_i$ is defined as:
\begin{equation}
\textit{C}_B(v_i) = \sum_{s\ne v\ne t\in V}^{} \frac{\sigma_{st}(v_i)}{\sigma_{st}}
\end{equation}
where $V$ denotes the nodes in the graph, $\sigma_{st}$ represents the number of shortest paths from node $s$ to node $t$, and $\sigma_{st}(v_i)$ is the number of shortest paths from node $s$ to node $t$ via $v_i$.

\textbf{PageRank centrality} measures the significance of a node in terms of the number of its connections~\cite{klicpera2018predict}. The intuition is that the more connected, the more important. The PageRank for a node $v_i$ is defined as:
\begin{equation}
\textit{PR}(v_i) = \alpha\sum_{v_j\in M_{v_j}}^{} \frac{\textit{PR}(v_j)}{\textit{L}(v_j)} +\frac{1-\alpha }{N} 
\end{equation}
where $M_{v_j}$ represents all neighbor nodes of $v_i$, $\textit{L}(v_j)$ is the number of neighbor nodes of $v_j$, $N$ is the number of nodes in the graph, and $\alpha$ denotes the dumping factor.

Finally, we are able to obtain the address graphs with abundant node features that include not only the semantic information of address transactions but also the augmented graph structural characteristics.


\subsection{Graph Representation Learning}

In this subsection, we use {{a graph neural network}} to learn the representation of the generated address graphs, producing the graph embeddings. It is challenging to choose an appropriate graph neural network for learning comprehensive and effective graph-level representation.

We adopt the graph feature network (GFN) to learn the graph representations. GFN can learn higher-order node representations and is extremely suited for address transaction graphs in terms of dealing with the features of nodes and the structure of the graph.
Graph representation learning module consists of three parts, \emph{namely} {graph feature augmentation}, {node representation learning}, and {graph representation readout}.

\textbf{Graph Feature Augmentation.}\quad Graph neural network enables the extraction of graph features, which essentially relies on the graph convolution process. The features of a node and its neighboring nodes in each graph convolution layer are aggregated to obtain a first-order node embedding. Consequently, the multi-scale graph convolution can achieve higher-order node embedding. However, too many layers of graph convolution might result in an overly smooth representation~\cite{li2019deepgcns, DBLP:journals/corr/abs-1909-12223}, which affects the final classification result.
Inspired by GFN, we represent the nodes in the graph using graph augmented features. We use the following two types of information to represent the nodes. One is the structural features of the graph, mainly the degree and centrality of the nodes. The other is the propagated features of the graph, which are extracted primarily from the higher-order adjacency matrix. The adjacency matrix is defined as follows:
\begin{equation}
\tilde{A} = \tilde{D}^{-\frac{1}{2}}(A+I)\tilde{D}^{-\frac{1}{2}} 
\end{equation}
where $I$ is the identity matrix, $\tilde{D}$ denotes the diagonal matrix, and $\tilde{A}$ is the normalized adjacency matrix. Then, we represent the graph feature extracted by GFN by
\begin{equation}
X^G = \left [ \textbf{d}, X, \tilde{A}^{1}X, \tilde{A}^{2}X, ..., \tilde{A}^{k}X  \right ]
\end{equation}
where $X^G$ is the augmented graph feature matrix, $\textbf{d}$ is the degree vector of each node, $X$ is the original node feature matrix and $\tilde{A}^{k}X$ is the $k$-th node feature matrix.

\textbf{Node Representation Learning.}\quad After the graph feature augmentation, we obtain a graph representation $X^G$ of the graph $G$. Then, we employ {{a multi-layer perceptron}} (MLP) to capture more effective graph features. By concatenating the MLP layer with the label of each graph, we can train the classification model to provide a more abstract representation of the nodes in a graph. This process can be expressed as follows:
\begin{equation}
\textit{GFN\_Classifier} = \sigma \left (W \sum \left ( X^G \right )  \right ) 
\end{equation}
where $\sigma()$ is {{a nonlinear activation function}} such as ReLU and $W$ is the trainable parameters of the neural network.

Compared with traditional graph convolutional networks~\cite{niepert2016learning}, GFN removes the adjacency matrix, and instead adopts neural networks to obtain higher-level node representations. As previously stated, the required information including higher-order node representation is obtained by feature expansion. By doing so, we can improve computation efficiency while maintaining more feature information.

\textbf{Graph Representation Readout.}\quad The readout function~\cite{lu2021learning} produces the graph representation by aggregating representation of all nodes in the graph. We employ the global pooling function SUM~\cite{bianchi2020spectral} to aggregate all the node embeddings in the graph to obtain the final graph embedding.
\begin{equation}
rep^G = \textit{SUM}(X^G_v), v\in V^G
\end{equation}
where the graph embedding $rep^G$ is the final representation of graph $G$.

\subsection{Address Classification}

So far, we have used graph neural networks to accomplish the mapping of address graphs to graph embeddings. Now, we consider representing the transaction features of a specific address as a list of embeddings with temporal features. Due to the distinct number of transactions at each address, the number of elements in the embedding list corresponding to each address is also different. The behavior of transactions in an address is stationary. It corresponds to a definite label such as mining pool, gambling, and exchange. Therefore, the goal of the address classification phase is to determine the label of an address by analyzing the embedding list of that address, which can be transformed into a problem of classifying a vector list of indefinite length. To achieve this, we use a LSTM (Long Short-Term Memory) to process the embedding list.

LSTM proposed by~\cite{hochreiter1997long} is often used to model processing contextual information due to its design features and is well suited for modeling temporal information. Particularly, transactions in the bitcoin network are typical unidirectional temporal data. The overall architecture of LSTM is a self-constantly cyclic structure, which can be expanded and represented as multiple identical units connected continuously. Compared with RNN~\cite{sherstinsky2020fundamentals}, LSTM outputs the memory of the current unit while outputting the hidden layer state and adding three gates.

\textbf{Forget gate} controls which information will be forgotten and which needs attention. The sigmoid function {{takes input $x_t$}} and hidden state $h_{t-1}$ and returns the values $f_t$ between 0 and 1, indicating whether the input should be forgotten.
\begin{equation}
f_t = \sigma (W_f\cdot [h_{t-1}, x_t] + b_f)
\end{equation}

\textbf{Input gate} controls how much information can be effectively input at any one time. Another sigmoid function accepts the current state $x_t$ and the previous hidden state $h_{t-1}$  and returns the value $i_t$ between 0 and 1.  Simultaneously, the tanh function receives $x_t$ and $h_{t-1}$ and returns a value between -1 and 1 for multiplication.
\begin{gather}
i_t = \sigma (W_i\cdot [h_{t-1}, x_t] + b_i) \\
\tilde{C_t}  = \tanh (W_C\cdot[h_{t-1}, x_t] + b_C)
\end{gather}

\textbf{Output gate} controls the output at the present time, which is the hidden state transmitted to the next unit. First, we need to update the cell state using the output of forget gate and input gate. By using the previous cell state $C_{t-1}$ multiplied with $f_t$ and $i_t$ multiplied with $\tilde{C_t}$, a new cell state $C_t$ can be generated. Following that, the third sigmoid function accepts previous hidden state $h_{t-1}$ and current state $x_t$, and then multiplies the result with $\tanh(C_t)$, yielding the hidden state of the current moment $h_t$. Finally, the new hidden state $h_t$ will be transferred to the next unit.
\begin{gather}
C_t = f_t*C_{t-1} + i_t * \tilde{C_t} \\
o_t = \sigma (W_o[h_{t-1}, x_t] + b_o) \\
h_t = o_t * \tanh (C_t)
\end{gather}

With LSTM, we can transfer the graph representation list of an address into a vector. To obtain the final classification result, we concatenate an MLP layer following the LSTM layer to classify the graph representation list. We are able to get the parameters associated with the model by training it to achieve address classification.

For address $ad_i$, we can obtain a temporal graph list $g_i = \{g_{i 1}, g_{i 2}, ..., g_{i k}\}$ via graph construction. {{After graph representation learning}}, all graphs in $g_i$ will be converted into the representation list $rep_i = \{rep_{i 1}, rep_{i 2}, ..., rep_{i k}\}$. Finally, we utilize the following formula to classify the $rep_i$ of $ad_i$.
\begin{equation}
pred_i = \textit{MLP}(\textit{LSTM}(rep_i))
\end{equation}
$pred_i$ is the {{predicted label}} of address $ad_i$, which represents the behavioral characteristics of the address.

{{
\textbf{Workflow of Our System.}\quad The anonymity of Bitcoin has given rise to a large number of underground banks (a type of Service). Such underground banks exist on the dark web and perform money laundering, which is not regulated by the financial system. Such underground banks are often very hidden and difficult to discover through manual investigation. Here, we take money laundering as an example to illustrate how our proposed system achieves malicious behavior detection.

First of all, BAClassifier automatically analyzes a large number of addresses to determine if the address belongs to an underground bank (\emph{i.e.,} money laundering behavior). Given an address as the input, BAClassifier will slice all transactions of this address in a chronological relationship and construct the address transaction graph for each slice. Thereafter, BAClassifier adopts single-transaction address compression and multi-transaction address compression to resize the scale of the original address graph, and further extract the structural information of the compressed graph. After that, BAClassifier feeds the graph to a graph neural network to learn a graph representation. Finally, the address classifier will learn the graph representation according to the temporal sequence of transaction slices, and predict whether the address is an underground bank. Notably, with the help of our system, we can also obtain characteristics of addresses that are related to the detected underground banks, so as to dig out more hidden addresses of underground banks.}}

\section{Experiment}
In this section, we empirically evaluate our proposed methods on large-scale real world bitcoin addresses. We seek to answer the following research questions:
\begin{itemize}
\item \textbf{RQ1:} How to construct a benchmark dataset for evaluating the performance of bitcoin address classification methods?
\item \textbf{RQ2:} Is our proposed model combination effective in learning graph representation and detecting address behaviors? 
\item \textbf{RQ3:} How are the \emph{precision}, \emph{recall}, and \emph{F1-score} performance of BAClassifier compared with state-of-the-art classifiers and classification models?
\item {{\textbf{RQ4:} What is the running overhead of each component in BAClassifier, namely address graph construction, graph representation learning, and address classification?}}
\end{itemize}
Next, we first present the experimental settings, followed by answering the above research questions one by one.


\subsection{Experimental Setting}
\paragraph{Implementation details} All experiments are conducted on a server equipped with two Intel Xeon E5-2680 v4 CPUs at 2.40GHz, 128G memory, and four Nvidia 2080Ti GPUs. All the modules in this work are implemented in Python programming language, and the neural network models are developed using the PyTorch framework.

\paragraph{Evaluation metrics} To measure the performance of a classification model, {{we adopt widely used metrics}}, namely \emph{precision}, \emph{recall}, and \emph{F1-score}. Precision is defined as a ratio of true positives and the total number of positives predicted by a tool. Recall is the fraction of the total number of relevant instances that are actually detected. F1-score is the harmonic average of precision and recall, and it is usually treated as the critical evaluation indicator for some classification missions.
\begin{gather}
\textit{Precision} = \frac{\textit{true\ positive}}{\textit{true\ positive} + \textit{false\ positive}} \\
\textit{Recall} = \frac{\textit{true\ positive}}{\textit{true\ positive} + \textit{false\ negative}} \\
\textit{F1-score} = 2\times \frac{\textit{Precision}\times \textit{Recall}}{\textit{Precision} + \textit{Recall}} 
\end{gather}

\subsection{RQ1: Dataset Construction}
To the best of our knowledge, there is still a lack of benchmark datasets for bitcoin address behavior classification. Indeed, it is non-trivial to obtain a high-quality dataset that contains labeled behaviors of the bitcoin address, attributing to the demand for qualified expertise. Motivated by this, we construct and release a benchmark dataset, which consists of over 2 million bitcoin addresses as well as their transactions. We develop an automated crawler tool and create the dataset by collecting bitcoin addresses from different sources: btc.com~\cite{Btc22}, walletexplorer.com~\cite{Walletexplorer22}, oxt.me~\cite{Oxt22}, and other websites. {{Note that all the bitcoin addresses are collected from trusted entities in the field. For example, the addresses of the exchange type are obtained from real-world exchanges, the addresses of the gambling type are disclosed by the gambling website, and the addresses of the mining type are provided by the mining pool service provider. Furthermore, we perform a manual secondary validation to ensure the quality of the collected data from open-sourced websites.}}

\textbf{Dataset Labeling.}\quad We manually label the bitcoin addresses in the dataset, \emph{i.e.,} what transaction behavior is being performed by a bitcoin address. We concentrate on four categories of behaviors, \emph{namely} exchange, mining, gambling, and service. Our annotations facilitate the evaluation of bitcoin address classification tools extensively.

\emph{(1) Exchange:} Exchanges hold this type of address, which consists of cold wallet addresses and hot wallet addresses. These addresses are often used by exchanges to manage funds and provide deposit and withdrawal services. 

\emph{(2) Mining:} {{This type of address is held by a mining pool}}. The mining pools receive and distribute mining rewards through this type of address, while the mining nodes receive their reward from the mining pools through this type of address.

\emph{(3) Gambling:} This type of address is held by gambling websites and gamblers. Gambling websites absorb and manage gambling funds through this class of addresses, while gamblers send and receive gambling funds through this class of addresses.

\begin{table}
 \centering
 \caption{The statistical analysis of our released dataset.}
 \label{tab:pagenum}
 \resizebox{0.23\textwidth}{!}{
 \begin{tabular}{cc}
  \toprule
  \textbf{Address Label} & \textbf{Number} \\
  \midrule
  Exchange  & 912,322\\
  Mining  & 133,119\\
  Gambling  & 377,559\\
  Service  & 715,657\\
  \midrule
  \textbf{Total}  & 2,138,657\\
  \bottomrule
 \end{tabular}
 }
\label{Statics}
\end{table}

\emph{(4) Service:} This type of address is held by websites that provide related services that provide related services, including wallet, coin mixer, dark web, and lending. {{Note that we do not care whether a certain behavior is anonymous. For example, for mixing services, we do not care where the money goes. We only focus on which addresses are used for money laundering and mixing services.}}

{{Labeling for bitcoin addresses is based on the characteristics of address transaction behaviors. Undoubtedly, it is difficult to define all kinds of transaction types. The reason why we select these four high-level categories is that they cover main bitcoin transaction behaviors, and there are obvious differences between these four types of behaviors, which can be regarded as good examples to prove the generality and scalability of our system. In future research, we will further refine the bitcoin address behaviors, such as exchange cold wallets, exchange hot wallets, decentralized exchange addresses, and so on.}}

\textbf{Dataset Statistic and Analysis.}\quad Table~\ref{Statics} showcases the dataset statistics. Specifically, 912,322 bitcoin addresses have executed the exchange operation. 133,119 bitcoin addresses have performed the mining behaviors. Around 377,559 bitcoin addresses have issued the gambling transactions. 715,657 bitcoin addresses provide other related services.

For the following experiments, we conducted random stratified sampling based on label types and selected around 10,000 addresses to participate in the training and testing of the model. We also partitioned over 10,000 addresses through random stratified sampling and selected 80\% of the addresses as the train set and 20\% of the addresses as the test set.

\renewcommand{\arraystretch}{1.1}
\begin{table}
\centering
\small
 \caption{ Performance comparison using between different graph representation models.}
\resizebox{0.49\textwidth}{!}{
\begin{tabular}{|c|c|c|c|c|}
  \hline
    \textbf{Methods} & \textbf{Model} & \textbf{Precision} & \textbf{Recall} & \textbf{F1-score} \\
  \hline
    \multirow{3}{*}{GNNs}
    & \textbf{GFN (ours)} & \textbf{0.9815} & \textbf{0.9725} & \textbf{0.9769} \\
      \cline{2-5}
    & Diffpool & 0.9218 & 0.9315 & 0.9299 \\
      \cline{2-5}
    & GCN & 0.9534 & 0.9461 & 0.9514 \\
  \hline
    \multirow{9}{*}{MLs}
    & LR & 0.2208 & 0.3477 & 0.2684 \\
      \cline{2-5}
    & MLP & 0.1011 & 0.2500 & 0.1440 \\
      \cline{2-5}
    & SVM & 0.8787 & 0.5503 & 0.5574 \\
      \cline{2-5}
    & Bernoulli NB & 0.5078 & 0.3434 & 0.3047 \\
      \cline{2-5}
    & Gaussian NB & 0.5342 & 0.4418 & 0.3999 \\
      \cline{2-5}
    & KNN & 0.8661 & 0.8553 & 0.8598 \\
      \cline{2-5}
    & Decision Tree & 0.9298 & 0.9178 & 0.9236 \\
      \cline{2-5}
    & GBDT~\cite{friedman2001greedy} & 0.9596 & 0.9575 & 0.9585 \\
      \cline{2-5}
    & XGBoost~\cite{chen2016xgboost} & 0.9340 & 0.9321 & 0.9329 \\
 \hline
 \end{tabular}}
 \label{GRMS}
 \end{table}

\subsection{RQ2: Model Selection}
In our framework, we propose a two-stage classification model, which contains a graph representation learning module and an address classification module. For graph representation learning, we transfer the address transactions into the address graph and choose graph feature network (GFN)~\cite{chen2019powerful} to learn the graph representation. For the address classification, we adopt the combination of LSTM+MLP. 
In this subsection, we compare our selected GFN and LSTM+MLP models with existing other popular models to illustrate the validity of the model selection in our proposed framework.

\subsubsection{Graph Representation Model}
In contrast to previous works~\cite{lee2020machine,zhang2018bitscope}, our study is the first to apply graph neural networks to bitcoin address behavior classification. To demonstrate the effectiveness of graph neural networks in processing address transactions, we compare graph neural networks with traditional machine learning methods (\emph{e.g.,} LR, SVM, MLP, GBDT, XGBoost). In addition, we also compare  GFN with other kinds of graph neural networks such as GCN, to verify that GFN outperforms other graph neural networks in processing the generated address transaction graphs.

\renewcommand{\arraystretch}{1.1}
\begin{table*}
\centering
\small
\caption{Performance comparison in terms of recall, precision, and F1-score. A total of six address classification models are investigated in the comparison. Weighted Avg represents the average results for all address classifications across the dataset.}
\resizebox{0.998\textwidth}{!}{
\begin{tabular}{|c|c|c|c|c|c|c|c|c|c|}
  \hline
    \textbf{Model} & \textbf{Type} & \textbf{Precision} & \textbf{Recall} & \textbf{F1-score} &
    \textbf{Model} & \textbf{Type} & \textbf{Precision} & \textbf{Recall} & \textbf{F1-score} \\
  \hline
    \multirow{5}{*}{\makecell[c]{\textbf{LSTM+MLP} \\ \textbf{(ours)}}}
    & Exchange & 0.9763 & 0.9891 & 0.9826 &
    \multirow{5}{*}{BiLSTM+MLP}
    & Exchange & 0.9603 & 0.9967 & 0.9782 \\
      \cline{2-5} \cline{7-10}
    & Mining & 0.9827 & 0.9660 & 0.9743 &
    & Mining & 0.9962 & 0.9887 & 0.9924 \\
      \cline{2-5} \cline{7-10}
    & Gambling & 0.9916 & 0.9869 & 0.9893 &
    & Gambling & 0.9940 & 0.9617 & 0.9776 \\
      \cline{2-5} \cline{7-10}
    & Service & 0.8966 & 0.8125 & 0.8525 &
    & Service & 0.9032 & 0.7778 & 0.8358 \\
      \cline{2-5} \cline{7-10}
    & Weighted Avg & 0.9618 & 0.9386 & \textbf{0.9497} &
    & Weighted Avg & 0.9634 & 0.9312 & 0.9460 \\
  \hline
    \multirow{5}{*}{Attention+MLP}
    & Exchange & 0.9935 & 0.9715 & 0.9824 &
    \multirow{5}{*}{SUM+MLP}
    & Exchange & 0.9785 & 0.9934 & 0.9859 \\
      \cline{2-5} \cline{7-10}
    & Mining & 0.9761 & 0.9371 & 0.9879 &
    & Mining & 0.9103 & 0.9864 & 0.9932 \\
      \cline{2-5} \cline{7-10}
    & Gambling & 0.9804 & 0.9942 & 0.9872 &
    & Gambling & 0.9905 & 0.9881 & 0.9893 \\
      \cline{2-5} \cline{7-10}
    & Service & 0.8205 & 0.8889 & 0.8533 &
    & Service & 0.9615 & 0.7353 & 0.8333 \\
      \cline{2-5} \cline{7-10}
    & Weighted Avg & 0.9426 & 0.9479 & 0.9452 &
    & Weighted Avg & 0.9652 & 0.9258 & 0.9450 \\
  \hline
    \multirow{5}{*}{AVG+MLP}
    & Exchange & 0.9859 & 0.9827 & 0.9843 &
    \multirow{5}{*}{MAX+MLP}
    & Exchange & 0.9910 & 0.9746 & 0.9827 \\
      \cline{2-5} \cline{7-10}
    & Mining & 1.0000 & 0.9840 & 0.9919 &
    & Mining & 0.9632 & 0.9968 & 0.9797 \\
      \cline{2-5} \cline{7-10}
    & Gambling & 0.9930 & 0.9850 & 0.9890 &
    & Gambling & 0.9858 & 0.9893 & 0.9875 \\
      \cline{2-5} \cline{7-10}
    & Service & 0.6981 & 0.9487 & 0.8043 &
    & Service & 0.8261 & 0.8636 & 0.8444 \\
      \cline{2-5} \cline{7-10}
    & Weighted Avg & 0.9193 & 0.9751 & 0.9424 &
    & Weighted Avg & 0.9415 & 0.9561 & 0.9486 \\
  \hline
\end{tabular}}
\label{ACMS}
\end{table*}

{{Unlike graph neural networks, traditional machine learning models cannot directly process the graph structural information~\cite{zhang2018bitscope,lee2020machine}. Therefore, for a feasible comparison, we transform graph features into feature inputs that match the form of machine learning models. Specifically, we first obtain the feature vector of each node (address) in the address transaction graph. Thereafter, we aggregate feature vectors of all input nodes and all output nodes of a target node, respectively. Finally, we generate the final feature input of a machine-learning model by concatenating the aggregated feature vector of input nodes, the feature vector of the target node, and the aggregated feature vector of output nodes.}}
Table~\ref{GRMS} summarizes the experimental results. From the table, we can observe that GFN obtains the best performance rating out of the compared 12 models. Specifically, the F1-score of GFN increases by about 2.5 percentage points compared to when GCN. The F1-score of GFN also increases by about 1.8 percentage points when compared to GBDT which usually achieves the best results in classification tasks.

\subsubsection{Address Classification Model}
After the graph representation learning, we merge the representations of multiple graphs corresponding to one address. There are two primary methods for training the graph representations. One is to use recurrent neural networks (\emph{e.g.,} LSTM), and the other utilizes pooling operations (\emph{e.g.,} Attention Pooling, Sum Pooling, Avg Pooling, Max Pooling). Due to the temporal relationship between different nodes in the graph, we consider processing the graph representation by employing LSTM which is connected with several MLP layers for the address classification. To confirm our model selections, we try to adopt different combinations for the experimental comparison.

Specifically, we adopt six different combinations for classifying the graph representations (\emph{i.e.,} outputting the address classification). Due to the excellent performance earned by GFN in graph representation learning, all the six types of combinations in this comparing experiment have achieved accurate classification results. By comparing the LSTM model to the bidirectional LSTM model, we found that the LSTM model achieved better results, confirming our conjecture that bitcoin transactions are precisely a forward dependency, with the output of the first transaction determining the input of the subsequent transaction while the subsequent transaction being unable to change the state of the earlier transactions. Additionally, we compare the LSTM method to the pooling operation, which is a frequently used method for aggregation. We compare the outcomes of Attention Pooling, Sum Pooling, Avg Pooling, and Max Pooling to the LSTM model. From Table~\ref{ACMS}, we can see that the results of LSTM model outperform all the pooling operations. We speculate the reason lies in that the pooling operation is unable to process the temporal features in the graph, and hence the F1-score of the classification is relatively lower.

\renewcommand{\arraystretch}{1.1}
\begin{table}
\centering
 \caption{{{Performance comparison between BAClassifier and other bitcoin address classifiers. Weighted Avg represents the average results for all address classifications across the dataset.}}}
 \resizebox{0.49\textwidth}{!}{
\begin{tabular}{|c|c|c|c|c|}
  \hline
    \textbf{Classifiers} & \textbf{Type} & \textbf{Precision} & \textbf{Recall} & \textbf{F1-score} \\
  \hline
    \multirow{5}{*}{BAClassifier}
    & Exchange & 0.9763 & 0.9891 & 0.9826 \\
      \cline{2-5}
    & Mining & 0.9827 & 0.9660 & 0.9743 \\
      \cline{2-5}
    & Gambling & 0.9916 & 0.9869 & 0.9893 \\
      \cline{2-5}
    & Service & 0.8966 & 0.8125 & 0.8525 \\
      \cline{2-5}
    & \textbf{Weighted Avg} & \textbf{0.9618} & \textbf{0.9386} & \textbf{0.9497} \\
  \hline
    \multirow{4}{*}{BitScope~\cite{zhang2018bitscope}}
    & Exchange & 0.7800 & 0.8900 & 0.8300 \\
      \cline{2-5}
    & Mining & 0.8700 & 0.7600 & 0.8100 \\
      \cline{2-5}
    & Gambling & 0.7600 & 0.6900 & 0.7200 \\
      \cline{2-5}
    & Service & 0.7700 & 0.7100 & 0.7400 \\
  \hline
    \multirow{4}{*}{\makecell[c]{Lee et al.~\cite{lee2020machine} \\ with \\ Random Forest}}
    & Exchange & 0.8200 & 0.7400 & 0.7800 \\
      \cline{2-5}
    & Mining & 0.8600 & 0.8500 & 0.8600 \\
      \cline{2-5}
    & Gambling & 0.7700 & 0.7700 & 0.7700 \\
      \cline{2-5}
    & Service & 0.7800 & 0.8600 & 0.8200 \\
  \hline
    \multirow{4}{*}{\makecell[c]{Lee et al.~\cite{lee2020machine} \\ with \\ ANN}}
    & Exchange & 0.6200 & 0.5200 & 0.5600 \\
      \cline{2-5}
    & Mining & 0.7700 & 0.5600 & 0.6500 \\
      \cline{2-5}
    & Gambling & 0.5100 & 0.4600 & 0.4800 \\
      \cline{2-5}
    & Service & 0.4500 & 0.4500 & 0.4500 \\
 \hline
 \end{tabular}}
 \label{MC}
 \end{table}

\subsection{RQ3: Performance Comparison}

In this subsection, we benchmark BAClassifier against existing bitcoin address classification tools to further illustrate the advancement of our proposed method. We compare with BitScope~\cite{zhang2018bitscope} and the machine learning-based classifier~\cite{lee2020machine} proposed by Lee \emph{et al.} BitScope classifies the bitcoin address with a layered approach and exploits the domain-specific structures in the bitcoin transaction network. Machine learning-based classifier proposed by Lee \emph{et al.} extracts 80 features from the bitcoin transactions and uses two different models (\emph{i.e.,} random forest and ANN) to classify the bitcoin address.

Table~\ref{MC} showcases the comparison results. Obviously, our method achieves significant improvement in the classification of each category of address behavior. For example, BAClassifier achieves a 98\% F1-score on the classification of exchange address, 15\%, 20\%, and 32\% higher than the other three compared methods, respectively. This can be attributed to the fact that previous methods have heavily relied on feature extraction to describe the transaction features and correlations of bitcoin addresses, which results in significant information loss, such as semantic information and temporal relationship. In contrast, our approach leverages the graph neural networks which are able to characterize the graph features, and uses the LSTM networks to model the temporal information between address transactions.

\begin{figure}
    \centering
    \includegraphics[width=0.5\linewidth]{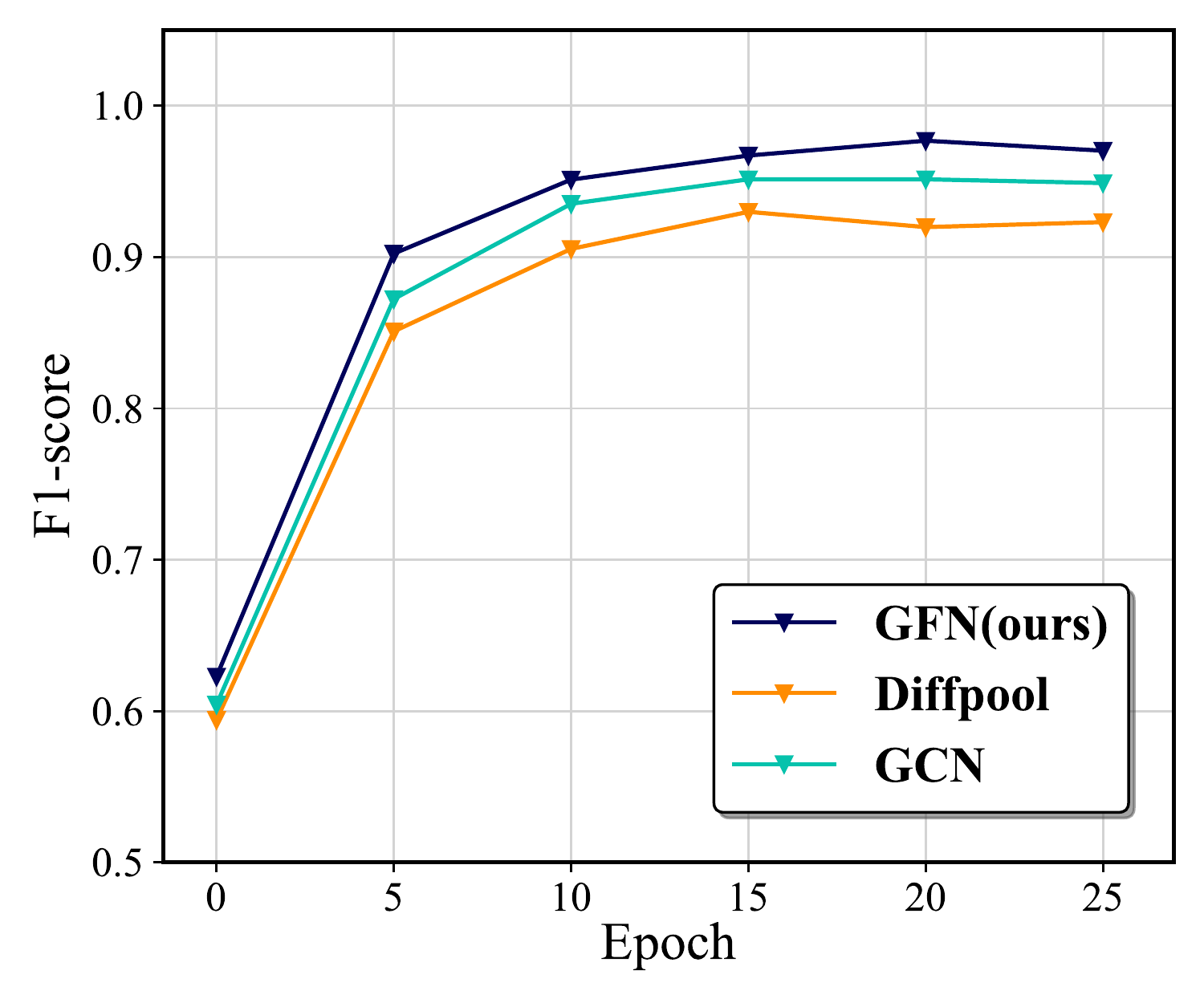}\includegraphics[width=0.5\linewidth]{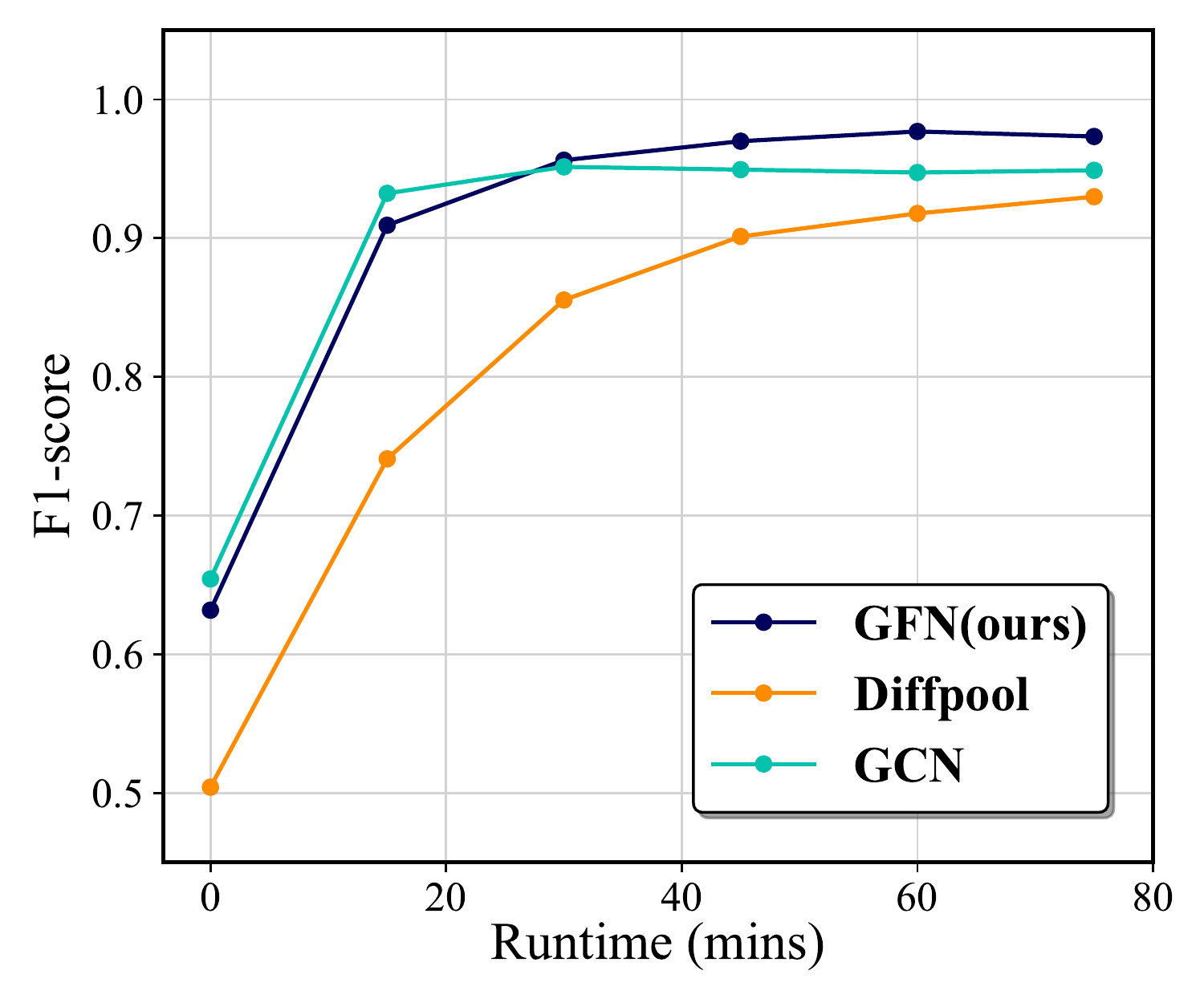}
    \caption{ {{Overhead evaluation for different graph neural networks in graph representation learning. The left figure shows the F1-score variation of the model for the same number of training epochs. The right figure shows the F1-score variation of the model with the same training runtime. }} }
    \label{fig:GRL}
\end{figure}

\begin{table}
 \centering
 \caption{ {{Runtime overhead and proportion for each stage of constructing address graphs.}} }
 \label{tab:graphconstructruntime}
 \resizebox{0.45\textwidth}{!}{
 \begin{tabular}{cccccc}
  \toprule
  \textbf{Metrics} & \textbf{Stage 1} & \textbf{Stage 2} & \textbf{Stage 3} & \textbf{Stage 4} & \textbf{Total} \\
  \midrule
  \textbf{CPU Runtime}  & 0.19s  & 0.63s  & 2.71s  &  0.81s & 4.34s\\
   \textbf{Ratio}  & 4.38\%  & 14.52\%  &  62.44\% & 18.66\%  & 100\%\\
  \bottomrule
 \end{tabular}
 }
\label{Static}
\end{table}

{{
\subsection{RQ4: Runtime Overhead Evaluation}

In this subsection, we further evaluate the runtime overhead of three components in \emph{BAClassifier}, respectively. 

\subsubsection{Address Graph Construction}
In this step, we translate transactions of bitcoin addresses into address graphs. To facilitate the runtime overhead evaluation of this procedure, we divide address graph construction into four stages, \emph{i.e.,} original graph extraction (Stage 1), single-transaction address compression (Stage 2), multi-transaction address compression (Stage 3), and graph structure augmentation (Stage 4). In Stage 1, our system extracts all original transactions of bitcoin addresses, slices these transactions, and generates original address transaction graphs. In Stage 2, we filter out all address nodes that have only one transaction associated with them, extract features of each address node, and aggregate features of these nodes. In Stage 3, we further screen out address nodes associated with multiple transactions and aggregate features of these nodes. In Stage 4, we extract the structural information of compressed address graphs and convey graphs to the upcoming graph representation learning component. It is worth mentioning that address graph construction is a CPU-dependent task, which can be processed in parallel using multiple processes. Here, we present the evaluation results of the running time (CPU time) for the address graph construction within a single CPU core. Specifically, we conduct experiments on our whole dataset and calculate the average processing time.

\begin{figure}
    \centering
    \includegraphics[width=0.5\linewidth]{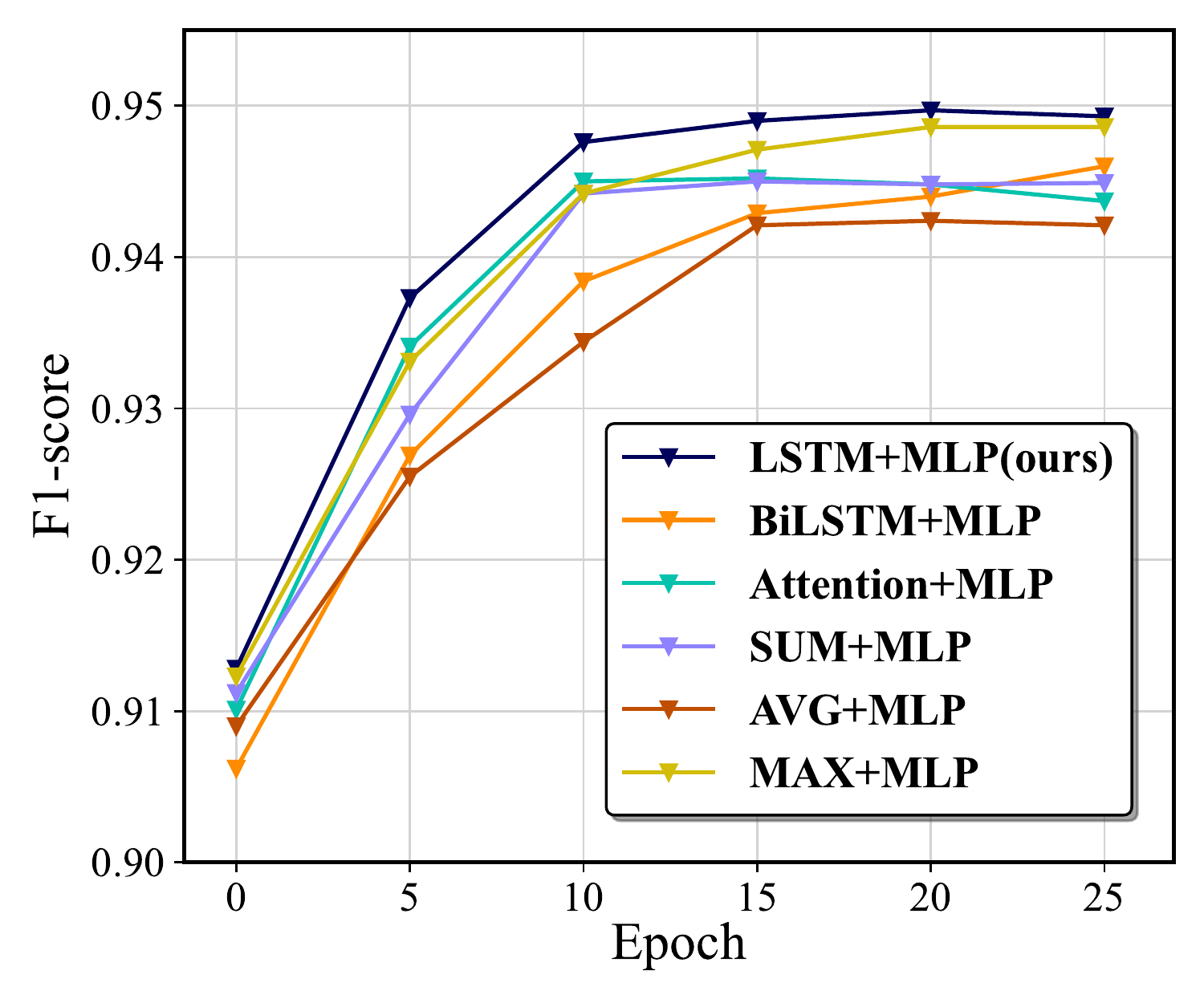}\includegraphics[width=0.5\linewidth]{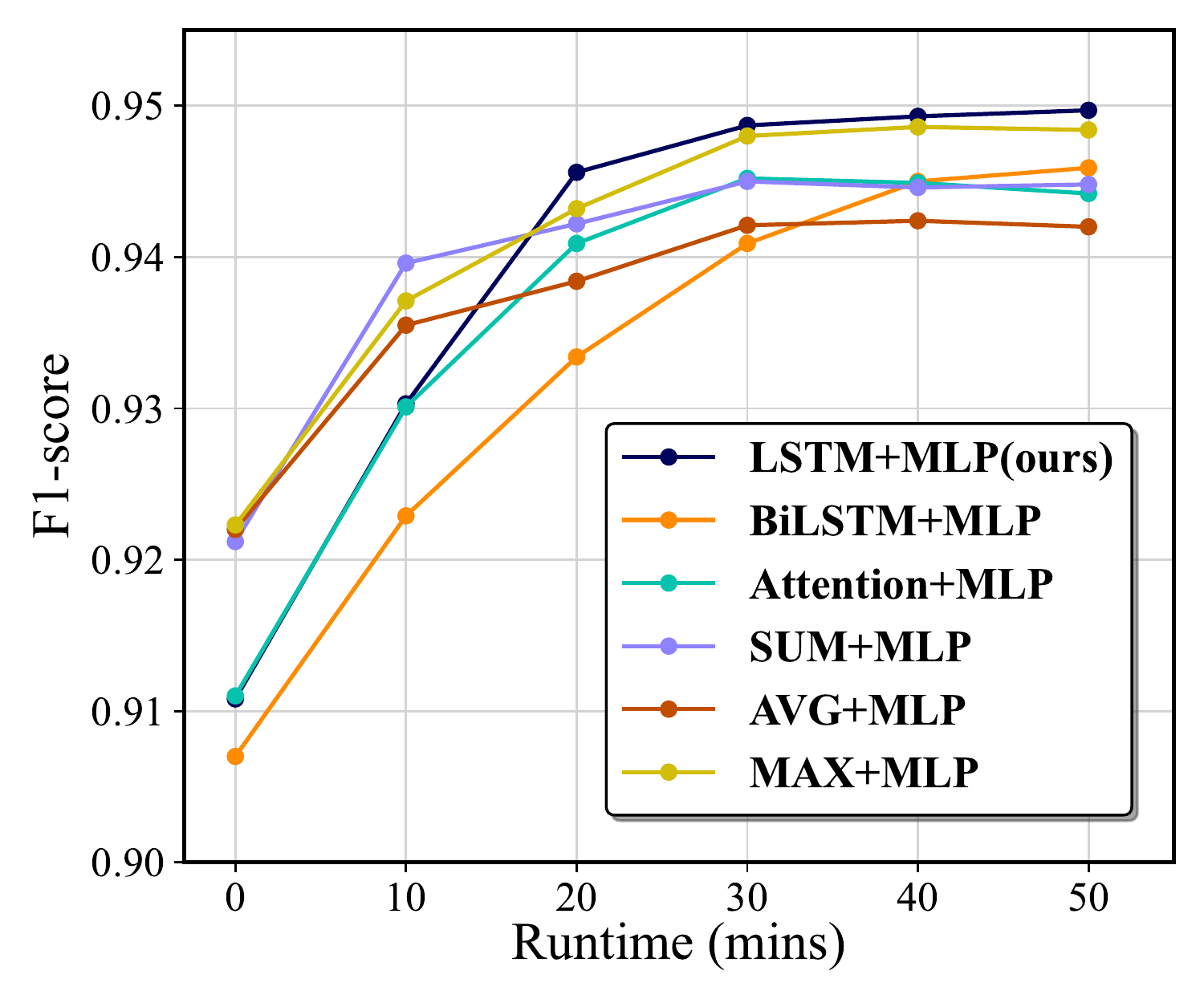}
    \caption{ {{Overhead evaluation for different address classification models. The left figure shows the F1-score variation of the model for the same number of training epochs. The right figure shows the F1-score variation of the model with the same training runtime.}} }
    \label{fig:AC}
\end{figure}

The quantitative results of each stage are summarized in Table~\ref{tab:graphconstructruntime}. From the table, we observe that the average single-core processing time for each address is 4.34s, of which the multi-transaction address compression (Stage 3) costs much time, accounting for 62.44\% of the total time. This may be attributed to the fact that this stage has to calculate the similarity between all different nodes.

\subsubsection{Graph Representation Learning}
Now, we conduct experiments to measure the overhead of the graph representation learning component. Fig.~\ref{fig:GRL} compares GFN to other graph models. From the figure, we observe that GFN is superior to others in terms of \emph{F1-score}. As shown on the left of Fig. 1, the number of epochs (within 20 epochs) required for GFN fitting is the same as in other models. Notably, the proposed model GFN consistently outperforms other approaches in each epoch. As shown in the right of Fig.~\ref{fig:GRL}, we can observe that GFN is superior to other methods in the same training time. For example, after 60 minutes of training, GFN gains a 97.69\% F1-score, 5.91\% and 2.96\% higher than GCN and Diffpool, respectively. The experimental results suggest that GFN achieves fast and effective graph feature extraction in the graph representation learning phase.

\subsubsection{Address Classification}
Similarly, in the address behavior classification step, we evaluate the F1-score of each classification model over training epoch and runtime, respectively. Fig.~\ref{fig:AC} visualizes the results, where the purple curve demonstrates the F1-score of our proposed LSTM+MLP over different epochs and training times. Clearly, the performance of LSTM+MLP is consistently better compared to other combinations across all epochs.

}}

\section{Conclusion and Future Work}
With the growing popularity of bitcoin, analyzing the behavior of bitcoin addresses has emerged as a significant research topic. The most effective way for behavioral analysis is to classify bitcoin addresses. In this work, we propose a novel framework \emph{BAClassifier} for bitcoin address classification. Specifically, we propose to transfer the bitcoin address transactions into the graph structures, and then utilize graph neural networks as the graph feature learning models. We evaluate the effectiveness of \emph{BAClassifier} over 2 million bitcoin addresses. Experimental results demonstrate that our proposed approach outperforms existing state-of-the-art address classifiers, which can accurately classify bitcoin addresses with a 95\% F1-score. Our implementation and datasets are released, hoping to push forward the boundary of this research direction.

In the future, we will further deepen this research from three aspects. To begin, we will improve the current dataset. The labeled dataset mainly classifies bitcoin addresses into four categories. We will expand the number of categories based on the address behavior. At the same time, the dataset only contains the type of address while lacking the entity information of the address (\emph{e.g.,} we are curious to know which exchange the address belongs to and whether it is from Coinbase, Binance, or another). Secondly, our model only utilizes the topology of the current node and other connected nodes when constructing the transaction graph, which does not take account into the label information of other nodes. In real-world scenarios, nodes of the same type often cluster together. The accuracy of the classification model can usually be improved by analyzing the types of connected nodes. Finally, we will investigate the characteristics of illegal behaviors on bitcoin such as money laundering and dark web transfer, in order to identify the addresses involved in illegal activities.

\section*{Acknowledgment}
This research is supported by the National Key R\&D Program of China (2021YFB2700500, 2021YFB2700501). The authors thank the anonymous reviewers for giving us a lot of valuable comments.

\bibliographystyle{IEEEtran}
\bibliography{conference_101719}

\end{document}